\documentclass[a4paper,12pt]{article}
\newcommand{\be}{\begin{equation}}
\newcommand{\ee}{\end{equation}}
\newcommand{\bea}{\begin{eqnarray}}

\newcommand{\eea}{\end{eqnarray}}

\begin{document}
\title{\bf {Bouncing cosmological solutions due to the self-gravitational corrections and their stability }}
\author{M. R. Setare, \footnote{E-mail: rezakord@ipm.ir},
\\{Physics Dept. Inst. for Studies in Theo. Physics and
Mathematics(IPM)}\\
{P. O. Box 19395-5531, Tehran, IRAN } }
\date{\small{}}
\maketitle
\begin{abstract}
In this paper we consider the bouncing braneworld scenario, in
which the bulk is given by a five-dimensional AdS black hole
spacetime with matter field confined in a $D_3$ brane. Exploiting
the CFT/FRW-cosmology relation, we consider the self-gravitational
corrections to the first Friedmann-like equation which is the
equation of the brane motion. The self-gravitational corrections
act as a source of stiff matter contrary to standard FRW
cosmology where the charge of the black hole plays this role.
Then, we study the stability of solutions with respect to
homogeneous and isotropic perturbations. Specifically, if we do
not consider the self-gravitational corrections, the AdS black
hole with zero ADM mass, and open horizon is an attractor, while,
if we consider the self-gravitational corrections, the AdS black
hole with zero ADM mass and flat horizon, is a repeller.
\end{abstract}
% \begin{document}
\newpage
% \vspace*{10mm}

 \section{Introduction}

Motivated by string/M theory, the AdS/CFT correspondence, and the
hierarchy problem of particle physics, braneworld models were
studied actively in recent years \cite{Hora96}-\cite{Rand99}. In
these models, our universe is realized as a boundary of a higher
dimensional spacetime. In particular, a well studied example is
when the bulk is an AdS space. In the cosmological context,
embedding of a four dimensional Friedmann-Robertson-Walker
universe was also
considered when the bulk is described by AdS or AdS black hole \cite%
{Nihe99,AdSbhworld}. In the latter case, the mass of the black hole
was found to effectively act as an energy density on the brane with
the same equation of state of radiation. Representing radiation as
conformal matter and exploiting AdS/CFT correspondence, the
Cardy-Verlinde formula \cite{Verl00} for the entropy was found for
the universe ( see \cite{Seta02}, for the entropy formula in
the case of dS black hole ).\\
In either of the above cases, however, the cosmological evolution
on the brane is modified at small scales. In particular, if the
bulk space is taken to be an AdS black hole {\it with charge}, the
universe can `bounce' \cite{PEL}. That is, the brane makes a
smooth transition from a contracting phase to an expanding phase.
From a four-dimensional point of view, singularity theorems
\cite{hawk} suggest that such a bounce cannot occur as long as
certain energy conditions apply. Hence, a key ingredient in
producing the bounce is the fact that the bulk geometry may
contribute a negative energy density to the effective
stress-energy on the brane \cite{negative}. At first sight these
bouncing braneworlds are quite remarkable, since they provide a
context in which the evolution evades any cosmological
singularities while the dynamics is still controlled by a simple
(orthodox) effective action. In particular, it seems that one can
perform reliable calculations without deliberation on the effects
of quantum gravity or the details of the ultimate underlying
theory. Hence, several authors \cite{others,higherd,kanti,mset}
have pursued further developments for these bouncing braneworlds.
However the authors of \cite{mh} have found that generically these
cosmologies are in fact singular. In particular, they  have shown
that a bouncing brane must cross the Cauchy horizon in the bulk
space. However, the latter surface is unstable when arbitrarily
small excitations are introduced in the bulk spacetime.\\
Black hole thermodynamic quantities depend on the Hawking
temperature via the usual thermodynamic relations. The Hawking
temperature undergoes corrections from many sources: the quantum
corrections \cite{kaul}- \cite{set6}, the self-gravitational
corrections \cite{{KKW},{corrections}}, and the corrections due to
the generalized uncertainty principle \cite{gupdas, set9}.
Concerning the quantum process called Hawking effect
\cite{hawking1} much work has been done using a fixed background
during the emission process. The idea of Keski-Vakkuri, Kraus and
Wilczek (KKW) \cite{KKW} is to view the black hole background as
dynamical by treating the Hawking radiation as a tunnelling
process. The energy conservation is the key to this description.
The total (ADM) mass is kept fixed while the mass of the black
hole under consideration decreases due to the emitted radiation.
The effect of this modification gives rise to additional terms in
the formulae concerning the known results for black holes
\cite{corrections}; a nonthermal partner to the thermal spectrum
of the Hawking radiation shows up.
\\In this paper we take into account corrections to the
entropy of the  five-dimensional Schwarzschild- anti de Sitter
black hole (abbreviated to $SAdS_5$ in the sequel) that arise due
to the self-gravitational effect. The self-gravitational
correction, acts as a source for stiff matter on the brane, whose
equation of state is simply given by the pressure being equal to
the energy density. Due to the self-gravitational corrections, a
bouncing universe could, arise \cite{mset}. Then we describe
solutions of the bouncing braneworld theory and also determine
their stability. To do this, we use a set of convenient
phase-space variables similar to those introduced in
\cite{CS,GE}. The critical points of the system of differential
equations in the space of these variables describe interesting
non-static solutions. A method for evaluating the eigenvalues of
the critical points of the Friedmann and Bianchi models was
introduced by Goliath and Ellis \cite{GE} and further used in the
analysis by Campos and Sopuerta \cite{CS} of the Randall Sundrum
braneworld theory. These latter authors gave a complete
description of stationary points in an appropriately chosen phase
space of the cosmological setup and investigated their stability
with respect to homogeneous and isotropic perturbations. The
authors worked in the frames of the Randall Sundrum braneworld
theory without the scalar-curvature term in the action for the
brane. Cosmological solutions and their stability with respect to
homogeneous and isotropic perturbations in the braneworld model
with the scalar-curvature term in the action for the brane have
further been studied by Iakubovskyi and Shtanov \cite{shta}.
\section{Self-Gravitational Corrections to FRW Equation}
In the asymptotic coordinates, the $SAdS_5$ black hole metric is
 \bea
ds^{2}=-F(r)dt^{2}+\frac{1}{F(r)}dr^{2}+r^{2} d\Omega_{(3)}^{2},
\label{metric1} \eea where \be
F(r)=1-\frac{\mu}{r^{2}}+r^{2},\label{fterm} \ee and we work in
units where the $AdS$ radius $l=1$. The parameter $\mu$ is
proportional to the ADM mass $M$ of the black hole.\\
We now consider a $4$-dimensional brane in the $SAdS_5$ black
hole background. This $4$-dimensional brane can be regarded as
the boundary of the $5$-dimensional $SAdS_{5}$ bulk background.
Let us first replace the radial coordinate $r$ with $a$ and so
the line element (\ref{metric1}) \bea
ds^{2}=-F(a)dt^{2}+\frac{1}{F(a)}da^{2}+a^{2}d\Omega_{(3)}^{2},
\label{met} \eea It was shown that by reduction from the $SAdS_5$
background (\ref{met}) and by imposing the condition \be
-F(a)\left(\frac{\partial t}{\partial \tau}\right)^{2}+
\frac{1}{F(a)}\left(\frac{\partial a}{\partial \tau}\right)^{2}=-1
\label{condition2} \ee where $\tau$ is a new time parameter, one
obtains an FRW metric for the $4$-dimensional timelike brane \be
ds_{(4)}^{2}=-d\tau^{2}+a^{2}(\tau)d\Omega_{(3)}^{2} \hspace{1ex}.
\ee Thus, the $4-$dimensional FRW equation describes the motion
of the brane universe in the $SAdS_5$ background. It is easy to
see that the matter on the brane can be regarded as radiation and
consequently, the field theory on the brane should be a CFT.

\par\noindent
Within the context of context the AdS/CFT correspondence,
Savonije and Verlinde studied the CFT/FRW-cosmology relation from
the Randall-Sundrum type braneworld perspective \cite{savonije}.
They showed that the entropy formulas of the CFT coincides with
the Friedmann equations when the brane crosses the black hole
horizon.

In the case of a $4$-dimensional timelike
 \be
ds_{(4)}^{2}=-d\tau^{2}+a^{2}(\tau)d\Omega_{(3)}^{2} \hspace{1ex},
\ee one of the identifications that supports the
CFT/FRW-cosmology relation  is \be
H^{2}=\left(\frac{2G_4}{V}\right)^{2}\mathcal{S}^{2} \label{huble}
\ee where $H$ is the Hubble parameter defined by
$H=\frac{1}{a}\frac{da}{d\tau}$ and V is the volume of the
$3$-sphere ($V=a^{3}V_{3}$), and $S$ is the entropy of the black
hole. The $4$-dimensional Newton constant $G_4$ is related to the
$5-$dimensional one $G_5$ by \be
 G_{4}=\frac{2}{l}G_{5}
 \hspace{1ex}.
\ee It was shown that at the moment that the $4$-dimensional
timelike brane crosses the cosmological horizon, i.e. when
$a=a_{b}$, the CFT entropy and the entropy of the $SAdS_5$ black
hole are identical. The modified Hubble equation, i.e. the first
Friedmann equation, takes the form \be
H^{2}=\frac{-1}{a_{b}^{2}}+\frac{8\pi G_{4}}{3}\rho - \frac{8\pi
G_{4}}{3}\left[\frac{4\pi
G_{4}}{3}\frac{1}{a_{b}^{2}V_{3}}\rho\right]\omega
\label{modHubble} \ee where the volume $V$ is given by
$a_{b}^{3}V_{3}$, $\rho$ is the energy density and $\omega$ is
the energy of an emitted particle from the black hole. At this
point it should be stresses that our analysis was up to now
restricted to the spatially flat ($k=+1$) spacelike brane.
\par
We will now extend the aforesaid analysis. We therefore consider
an arbitrary scale factor $a$ and include a general $k$ taking
values ${+1,0,-1}$ in order to describe, respectively, the
elliptic, flat, and hyperbolic horizon geometry of the $SAdS_5$
bulk black hole. The modified Hubble equation is now given by \be
H^{2}=\frac{-k}{a^{2}}+\frac{8\pi G_{4}}{3}\rho - \frac{8\pi
G_{4}}{3}\left[\frac{4\pi
G_{4}}{3}\frac{1}{a^{2}V_{3}}\rho\right]\omega \label{gmodHubble}
\ee where the volume $V$ is now given by $a^{3}V_{3}$   since all
quantities that appear in equation (\ref{gmodHubble}) are defined
for an arbitrary scale factor $a$.

The first term in the right-hand side of equation
(\ref{gmodHubble}) represents the curvature contribution to the
brane motion. The second term can be regarded as the contribution
from the radiation and  it redshifts as $a^{-4}$. The last term
in the right-hand side of equation (\ref{gmodHubble}) is the
self-gravitational correction to the motion of $4$-dimensional
timelike brane moving in the $5$-dimensional Schwarzschild-anti de
Sitter bulk background. Since this term goes like $a^{-6}$, it is
obvious that it is dominant at early times of the brane evolution
while at late times the second term, i.e. the radiative matter
term, dominates and thus the last term can be neglected. The sign
of last term is opposite with respect to the standard situation,
one may expect that this sign difference could have interesting
cosmological consequences. Indeed, we will see that it is crucial
in allowing a nonsingular transition between a contracting and an
expanding evolution of the scale factor $a$.

\section{Stability of the Bouncing Solutions}
In this section, we describe the bouncing  solutions of the
braneworld theory under investigation and also determine their
stability. To do this, we use a set of convenient phase-space
variables similar to those introduced in \cite{CS,GE}. At first
we rewrite (\ref{gmodHubble})as \bea
H^{2}=\frac{-k}{a^{2}}+\frac{\varepsilon_{3}M}{a^{4}}-
\frac{\varepsilon_{3}^{2}M\omega}{2a^{6}}
 \label{stiff} \eea
 where
 \be \varepsilon_{3}=\frac{16\pi G_5}{3V_3}
\ee
 Now, we introduce the notation
similar to those of \cite{CS}
\begin{equation}\label{omeq}
\Omega_{k}=\frac{-k}{a^2H^2}=\frac{-k}{\dot{a}^{2}}, \hspace{1cm}
\Omega_{M}=\frac{\varepsilon_3 M}{a^4H^2},\hspace{1cm}
\Omega_{\omega}=\frac{-\varepsilon_{3}^{2}M \omega}{2a^6H^2}.
\end{equation}
and work in the $3$-dimensional $\Omega$-space
$(\Omega_{k},\Omega_{M},\Omega_{\omega})$. In this space, the
$\Omega$ parameters are not independent since the Friedmann
equation (\ref{stiff}) reads
\begin{equation}\label{constraint}
\Omega_{k}+\Omega_{M}+\Omega_{\omega} =1\, .
\end{equation}
Since these terms are non-negative they must belong to the
interval [0, 1] and hence, the variables $\Omega = (\Omega_{k},
\Omega_{M}, \Omega_{\omega} )$ define a compact state space.
Introducing the primed time derivative
\begin{equation}
' = \frac{1}{H} \frac{d}{dt} \, ,
\end{equation} one obtains the
system of first-order differential equations \cite{CS}
\begin{equation}\begin{array}{l}\label{omemgaeq}
\Omega_{k}'=2q\Omega_{k}\, , \smallskip \\
\Omega_{M}'=2(q-1)\Omega_{M}\, , \smallskip \\
\Omega_{\omega}' = 2 (q - 2) \Omega_\omega \, ,
\end{array}
\end{equation}
where
\begin{equation}\label{qeq}
q=\frac{-1}{H^{2}}\frac{\ddot{a}}{a}=\Omega_{M}+2\Omega_{\omega}.
\end{equation}
 The behavior of this system of equations in the neighborhood
of its stationary point is determined by the corresponding matrix
of its linearization. The real parts of its eigenvalues tell us
whether the corresponding cosmological solution is stable or
unstable with respect to the homogeneous perturbations
\cite{shta}.
 To begin with, we have to find
the critical points of this dynamical system, which can be
written in vector form as follows
\begin{equation}
 \Omega' =f(\Omega),
\end{equation}
where $f$ can be extracted from(\ref{omemgaeq}). The critical
points, $\Omega^\ast$, namely the points at which the system will
stay provided it is initially at there, are given by the condition
\begin{equation}
 f(\Omega^{\ast})=0.
\end{equation}
Their dynamical character is determined by the eigenvalues of the
matrix
\begin{equation}
 \frac{\partial f}{\partial\Omega}|_{\Omega=\Omega^{\ast}}.
\end{equation}
 If the real part of
the eigenvalues of a critical point is not zero, the point is said
to be {\em hyperbolic} \cite{CS}. In this case, the dynamical
character of the critical point is determined by the sign of the
real part of the eigenvalues:  If all of them are positive, the
point is said to be a {\em repeller}, because arbitrarily small
deviations from this point will move the system away from this
state.  If all of them are negative the point is called an {\em
attractor} because if we move the system slightly from this point
in an arbitrary way, it will return to it. Otherwise, we say the
critical point is a {\em saddle} point. \\We construct our models
as follows:
\\
\\
 \noindent {\bf (1)} \ The model $k$ , or
$( \Omega_{k}, \Omega_{M}, \Omega_{\omega}) = (1,0,0)$. We have
\begin{equation}
q=0\, ,
\end{equation}
and the eigenvalues are
\begin{equation}
\lambda_{M}=-2\, , \quad \lambda_{\omega}=-4 \, .
\end{equation}

 \noindent {\bf (2)} \ The model $M$ , or
$( \Omega_{k}, \Omega_{M}, \Omega_{\omega}) = (0,1,0)$. We have
\begin{equation}
q=1\, ,
\end{equation}
and the eigenvalues are
\begin{equation}
\lambda_{k}=2\, , \quad \lambda_{\omega}=-2 \, .
\end{equation}

 \noindent {\bf (3)} \ The model $\omega$ , or
$( \Omega_{k}, \Omega_{M}, \Omega_{\omega}) = (0,0,1)$. We have
\begin{equation}
q=2\, ,
\end{equation}
and the eigenvalues are
\begin{equation}
\lambda_{k}=4\, , \quad \lambda_{M}=2 .
\end{equation}
The dynamical system (\ref{omemgaeq}) has three hyperbolic
critical
points as follows: \\
\\
i) The model $k \:(k=-1)$,
$$
M=\omega=0, \:\: a(t)=t,
$$
with the critical point of an {\it attractor} type.\\
\\
ii) The model $M$,
$$
k=\omega=0, \:\: a(t)=(M\varepsilon_{3})^{1/4}\sqrt{2t},
$$
with the critical point of a {\it saddle point} type. \\
\\
iii) The model $\omega$,
$$
k=M=0, \:\: a(t)=(\frac{3\sqrt{3}}{4}\varepsilon_{3}\omega
t)^{1/3},
$$
with the critical point of a {\it repeller} type.

\section{Conclusion}
In this paper we have considered  a four-dimensional timelike
brane with non-zero energy density as the boundary of the
$SAdS_{5}$ bulk background. Exploiting the CFT/FRW-cosmology
relation, we have considered the self-gravitational corrections to
the first Friedmann-like equation which is the equation of the
brane motion. The additional term that arises due to the
semiclassical analysis, can be viewed as stiff matter where the
self-gravitational corrections act as the source for it. This
result is contrary to standard analysis that regards the charge of
$SAdS_{5}$ bulk black hole as the source for stiff matter. Then,
we have studied bouncing cosmological solutions and their
stability with respect to homogeneous and isotropic perturbations
in a braneworld theory.  The effects of the self-gravitational
corrections  of five-dimensional black hole in the bulk have been
considered. By including this effect in the analysis we have
obtained three models with the critical points of an {\it
attractor}, a {\it saddle point} and a {\it repeller},
respectively, and constructed the complete state space for these
cosmological models.\\
Also one can consider the situation as the present paper with
logarithmic corrections in $SAdS_{5}$  or in $SdS_{5}$ bulk
 backgrounds \cite{serg}, we hope to come back at future to this
important problem.

\end{document}